# METIS: The Mid-infrared ELT Imager and Spectrograph

Bernhard Brandl[1,2]
Felix Bettonvil[1]
Roy van Boekel[3]
Adrian Glauser[4]
Sascha Quanz[4]
Olivier Absil[5]
António Amorim[6]
Markus Feldt[3]
Alistair Glasse[7]
Manuel Güdel[8]
Paul Ho[9]
Lucas Labadie[10]
Michael Meyer[11]
Eric Pantin[12]
Hans van Winckel[13]
and the METIS Consortium[a]

[1] Leiden University, the Netherlands
[2] Faculty of Aerospace Engineering, TU Delft, the Netherlands
[3] Max Planck Institute for Astronomy, Heidelberg, Germany
[4] Department of Physics, ETH Zürich, Switzerland
[5] STAR Institute, Université de Liège, Belgium
[6] Faculdade de Ciencias da Universidade de Lisboa, Portugal
[7] UK Astronomy Technology Centre, Edinburgh, UK
[8] Department of Astrophysics, University of Vienna, Austria
[9] Institute of Astronomy and Astrophysics, Academia Sinica, Taipei, Taiwan
[10] I. Physikalisches Institut, Universität zu Köln, Germany
[11] Department of Astronomy, University of Michigan, Ann Arbor, USA
[12] CEA Saclay, IRFU, Gif-sur-Yvette, France
[13] Instituut voor Sterrenkunde, KU Leuven, Belgium

The Mid-infrared ELT Imager and Spectrograph (METIS) will provide the Extremely Large Telescope (ELT) with a unique window to the thermal- and mid-infrared (3–13 μm). Its single-conjugate adaptive optics (SCAO) system will enable high contrast imaging and integral field unit (IFU) spectroscopy ($R \sim 100\,000$) at the diffraction limit of the ELT. This article describes the science drivers, conceptual design, observing modes, and expected performance of METIS.

## Introduction

METIS will be a versatile instrument serving a wide range of science applications that target the cool and dusty Universe. Its main science drivers are studies of exoplanets and proto-planetary discs. To that end, METIS will offer a combination of high angular resolution — six times that of the James Webb Space Telescope (JWST) — high-contrast imaging (HCI) with coronagraphy, and high spectral resolution (up to $R \sim 100\,000$). Thanks to the ELT's 39-m aperture, METIS will have a sensitivity to spectrally unresolved emission lines similar to that of the JWST, and a point source sensitivity to continuum emission similar to that of Spitzer-IRAC. In terms of instruments at ESO's Very Large Telescope (VLT), METIS can be thought of as a combination of the Enhanced Resolution Imager and Spectrograph (ERIS), the Cryogenic high-resolution InfraRed Echelle Spectrograph (CRIRES), and the VLT Imager and Spectrometer for mid-InfraRed (VISIR), but enhanced with the spatial resolution and collecting area of the 39-m ELT aperture.

METIS is being designed and built by a consortium of 12 partner institutes from 10 countries[a]. The Principal Investigator is Bernhard Brandl and the project office is located at Leiden University. More information on the project, as well as the full list with names of the METIS team members, can be found on the METIS website[1].

While the origin of METIS dates back to its Phase A study in 2008/09, the instrument concept has evolved quite considerably since then; most notably streamlining the instrument modes and strengthening its high-contrast imaging performance. Phase B started in September 2015 and was concluded with a successful Preliminary Design Review in May 2019, aiming for a Final Design Review in 2022.

## METIS science

METIS will address a broad range of science cases. These include investigations of Solar System objects, young stellar clusters and massive star formation, active galactic nuclei (AGN), evolved stars, the centre of our Milky Way, and star formation in other galaxies. On top of that, there are two science areas that have been driving the key instrument requirements: circumstellar discs and extrasolar planets.

METIS will provide access to those regions of protoplanetary discs where the bulk of terrestrial planet formation takes place and will extend the existing studies of disc kinematics and (molecular) composition with the Atacama Large Millimeter/submillimeter Array (ALMA) toward the inner regions of the disc. Two key aspects will be the distribution of water and organic molecules in the innermost disc regions (for example, Pascucci et al., 2013; Banzatti et al., 2017) and the use of isotopologue abundances (of CO, for example) to investigate the mixing of material within a large number of discs (for example, Smith et al., 2009; Brown et al., 2013). Significantly, METIS will also be able to identify signatures of ongoing planet formation and even detect directly young forming planets embedded within discs from kinematical imprints (see, for example, Teague et al., 2018 and Pinte et al., 2018 for studies using ALMA). It is these applications that drive the requirements for the high-dispersion IFU mode in the $L$ and $M$ bands. For the first time, disc gaps and substructures, frequently detected with ALMA at (sub)millimetre wavelengths (for example, Andrews et al., 2018) and with the Spectro-Polarimetric High-contrast Exoplanet Research instrument (SPHERE) at near-infrared wavelengths in scattered light (for example, Avenhaus et al., 2018), can be revealed with comparable spatial resolution in the mid-infrared.

Truly new science is expected from the direct detection of extrasolar planets, for example, by constraining the luminosities of gas giant exoplanets that have an empirical mass estimate from radial velocity measurements (or, soon, from GAIA astrometry). The high-dispersion IFU mode in the $L$ and $M$ bands will allow the exoplanet community to expand significantly upon the pioneering work with CRIRES (and soon CRIRES+) to detect and characterise the atmospheric composition and dynamics of hot — and with METIS also warm — exoplanets (for example, Brogi et al., 2016; Birkby et al., 2017). In addition, METIS will be able to investigate less massive, Saturn- and Neptune-like objects, and not only the Jupiter-class objects accessible today. Finally, it is expected that METIS will take the first steps towards the direct detection and characterisation of nearby temperate terrestrial exoplanets (see Figure 1). Following up on the New Earths in the Alpha Centauri Region (NEAR) experiment (Kasper at al., 2019) on the VLT, METIS will be able to test observationally whether our nearest neighbours harbour rocky worlds (cf. Quanz et al., 2015). The nearest star known to host a rocky, temperate planet is Proxima Centauri (Anglada-Escudé et al., 2016) and METIS will be able to detect the planet directly and — possibly — probe its atmospheric composition (see Figure 1).

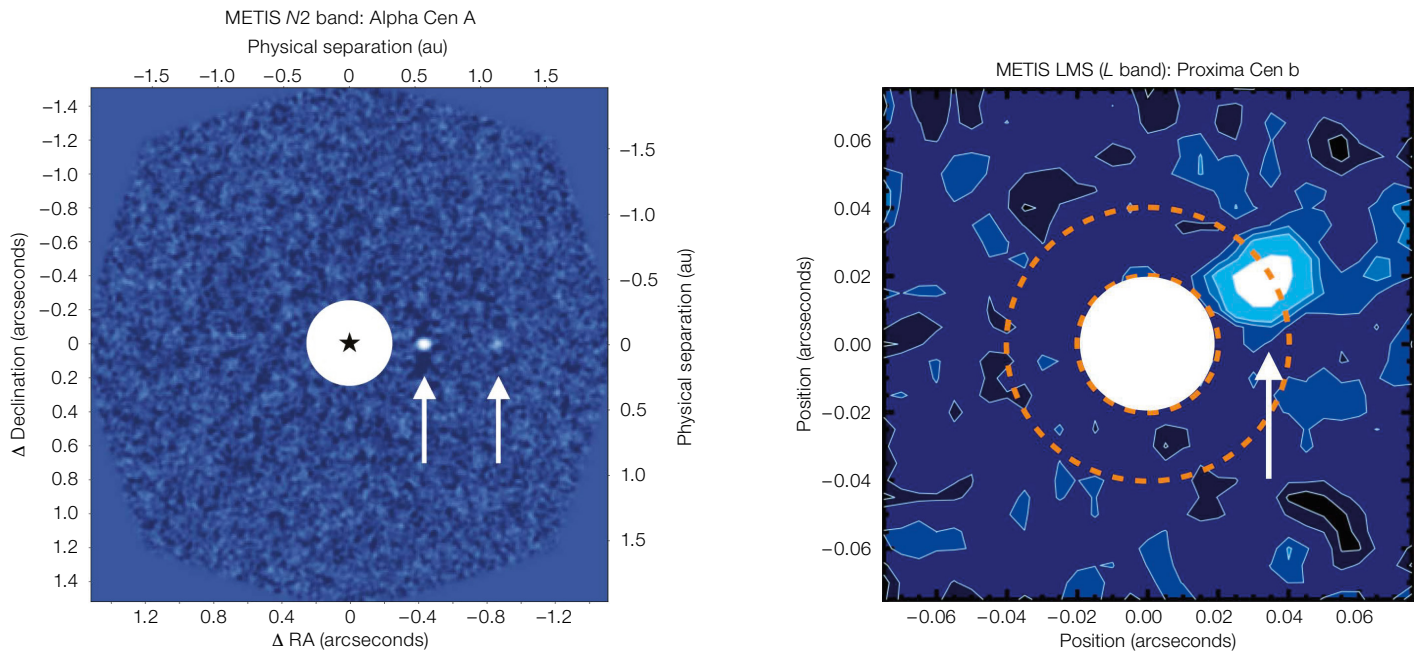

Figure 1. Left: Feasibility study to investigate if METIS can detect an Earth twin around Alpha Cen A at quadrature. These simulations are carried out with the METIS end-to-end high-contrast simulator and assume 5 hr of on-source time. In this case, two Earth-like (i.e., Earth radius and albedo) planets were inserted, one at 1.1 astronomical units (au) and one at 0.55 au (white arrows). Both planets are detected, with a signal-to-noise ratio of ~ 6 and ~ 10, respectively. Given the higher luminosity of the star, an Earth twin (i.e., same size and emission spectrum) would be located at ~ 1.1 au around Alpha Cen A. Right: Simulations of the METIS IFU performance on the Proxima system at 3.8 μm. Assuming a 1.1 $R_\oplus$ planet radius, a Bond albedo of 0.3, 50% illumination, an achieved coronagraph-aided contrast of 1:500 at 2 $\lambda/D$, and 10 hours of observing time, the planet is clearly detected in reflected light.

A lot of effort is currently being devoted to expanding the capabilities of the METIS instrument simulator to assess the feasibility of (new) science cases and the required observational effort. The community is warmly invited to download, test and use SimMETIS[2], provide critical feedback and expand the suite of exciting science cases addressable with METIS.

### Instrument concept

#### Main challenges

While any instrument on the ELT has to face numerous challenges, the two specific main challenges for METIS are the required high optical performance and the optimal control of the thermal background.

The METIS optical system is required to provide diffraction-limited performance in the *L*, *M* and *N* bands, corresponding to point spread functions ranging from 0.017 arcseconds FWHM at 3 μm to 0.070 arcseconds at 13 μm. METIS is equipped with a pyramid wavefront sensor inside the cryostat, operating in the *H* and *K* bands as part of the SCAO system which controls the ELT's adaptive mirrors M4 and M5. The AO guide star can be picked up anywhere within a circular field of view (FoV) with a diameter of 27 arcseconds, centred on the optical axis. In most cases the "guide star" will be the science target itself. To enhance the visibility of faint sources near bright stars, METIS uses various coronagraphic masks (see below), which require accurate and stable pupil alignment.

As with any ground-based thermal-infrared instrument, METIS needs to reduce the thermal emission from the atmosphere, the telescope mirrors and the telescope spiders to its fundamental limit. METIS is expected to operate in pupil-tracking mode to block the emission from the spiders effectively. Further background subtraction is accomplished via a combination of fast (~ 1 Hz) chopping offsets by a cold, fast, internal beam-chopping mirror, which enables more complex chopping patterns, and slow (~ 1 per min) nodding offsets by the telescope.

#### Conceptual design

The optical overview of METIS is shown in Figure 2. METIS consists of two science modules: a diffraction-limited imager with two wavelength channels, one for the *LM* band and one for the *N* band; and an IFU-fed, diffraction-limited, high-resolution *LM*-band spectrograph.

These science modules are attached to the Common Fore-Optics (CFO), which consists of two re-imagers that prepare the beam such that it arrives stabilised at the science modules. For this purpose, the CFO includes an atmospheric dispersion corrector for two fixed zenith angles, a derotator to stabilise the field or the pupil orientation, a pupil stabilisation mirror, a beam chopper for background reference measurements, and several pupil- and focal-plane wheels which host coronagraphs, slits and field masks. After the first re-imager, the light is spectrally split by a dichroic to feed the cryogenic AO wavefront sensor. A pick-off element can be inserted to guide the light into the *LM*-spectrometer.

The imager includes a common collimator, after which a dichroic element splits the light into the two wavelength channels where a set of filters, pupil masks and grisms can be inserted into the beam. A three-mirror-anastigmat camera focuses the light onto the science detectors: a Teledyne HAWAII-2RG in the *L* and *M* bands, and a Teledyne GeoSnap in the *N* band. The *LM*-spectrometer optically rearranges the field by means of a mirror slicer, followed by the pre-dispersion prism. The user can select between the full IFU field or a subset of slits with an extended wavelength coverage. After the high-resolution immersed grating the light is focused onto a 2 × 2 mosaic of Teledyne HAWAII-2RG detectors.

All optics are cryogenically cooled to ~ 70 K with the exception of the imager, which is at ~ 40 K, and the detectors, which are at 30–40 K. A cryostat provides the cryo-vacuum environment using three 2-stage pulse-tube coolers and liquid nitrogen for its radiation shields. The cryostat is supported by a support structure (Figure 3). All 22 cryo-mechanisms, heaters and sensors and all elements of the warm calibration unit are controlled by the instrument control system. A warm calibration unit on top of the cryostat provides calibration and alignment tools for daytime calibration and for the instrument integration and test phase.

## Observing modes

METIS offers five main observing modes, which are described in more detail below. These observing modes can be used in 24 different instrument configurations, not including the different science and neutral density filters. The online METIS app[3] offers a simple and interactive illustration of the METIS configuration for each observing mode.

Figure 2. Optical Layout of METIS.

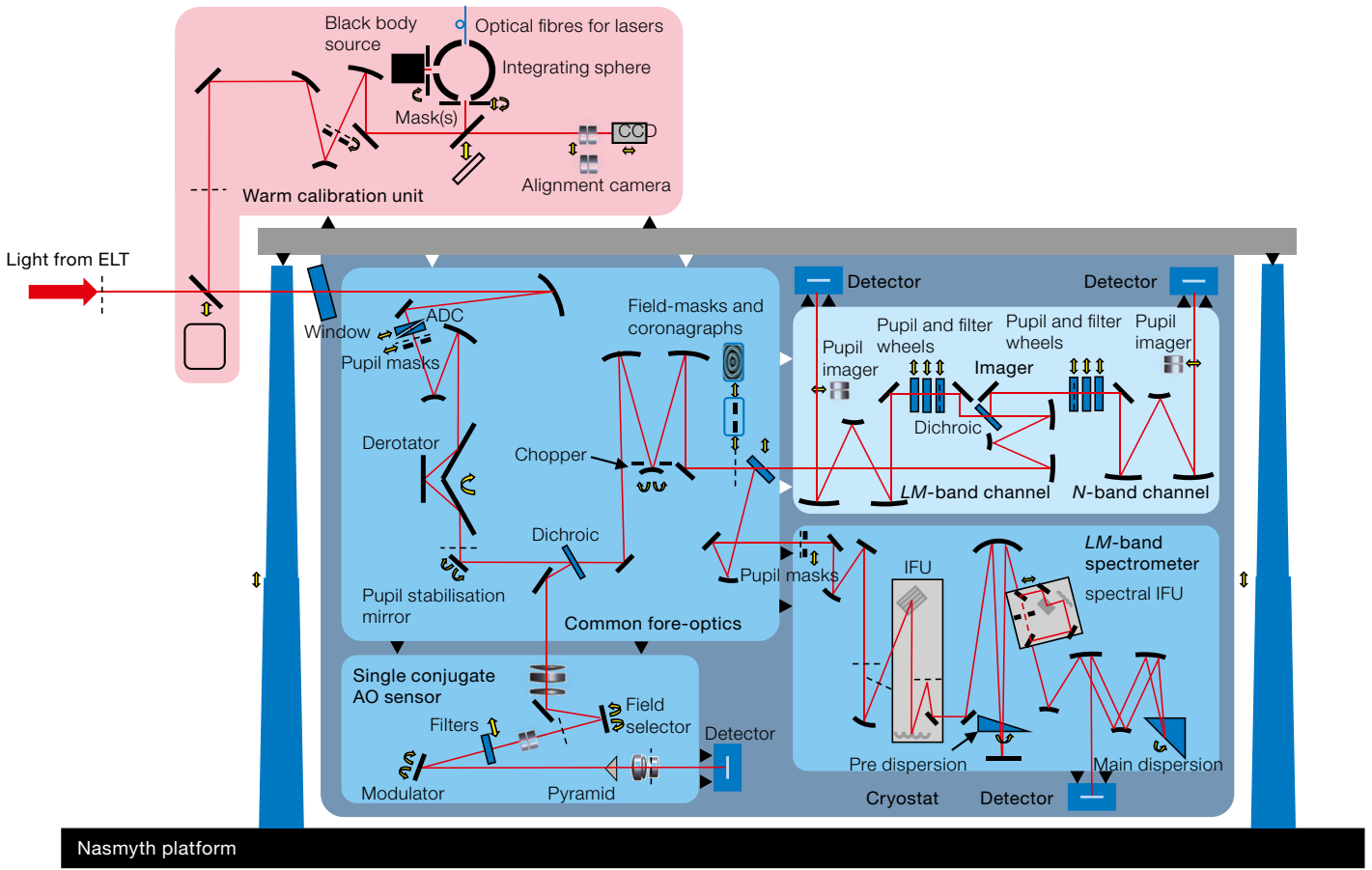

| | Filter | $\lambda_c$ | $\Delta\lambda$ | Point source | Surface brightness |
|---|---|---|---|---|---|
| Units | | μm | μm | μJy (5σ in 1 hr) | mJy arcsec$^{-2}$ (5σ in 1 hr) |
| Continuum imaging | $L'$ | 3.79 | 0.63 | 0.48 | 1.0 |
| | short-$L$ | 3.31 | 0.43 | 0.36 | 1.0 |
| | HCI-$L$ short | 3.60 | 0.22 | 0.53 | 1.2 |
| | HCI-$L$ long | 3.82 | 0.27 | 0.73 | 1.5 |
| | $M'$ | 4.80 | 0.60 | 3.8 | 5.0 |
| | $N1$ | 8.65 | 1.16 | 25 | 10 |
| | $N2$ | 11.20 | 2.36 | 30 | 7.2 |
| Spectral feature imaging | $H_2O$-ice | 3.10 | 0.22 | 0.35 | 1.1 |
| | PAH 3.3 | 3.30 | 0.07 | 1.3 | 3.5 |
| | Br-alpha | 4.05 | 0.03 | 4.4 | 8.0 |
| | $CO_{(1-0)}$/ice | 4.66 | 0.22 | 4.4 | 6.1 |
| | PAH 8.6 | 8.60 | 0.45 | 39 | 16 |
| | PAH 11.25 | 11.20 | 0.35 | 75 | 18 |
| | [S IV] | 12.82 | 0.23 | 89 | 24 |
| | [Ne II] | 10.50 | 0.19 | 170 | 31 |
| Longslit spectroscopy | $L$ (R ~1400) | 3.53 | | 5–9 × 10$^{-18}$ erg s$^{-1}$ cm$^{-2}$ | |
| | $M$ (R ~ 1900) | 4.9 | | 4 × 10$^{-17}$ erg s$^{-1}$ cm$^{-2}$ | |
| | $N$ (R ~ 400) | 10.50 | | 4 × 10$^{-16}$ erg s$^{-1}$ cm$^{-2}$ | |
| IFU spectroscopy | $L$ (R ~ 100 000) | 3.8 | | 7–20 × 10$^{-19}$ erg s$^{-1}$ cm$^{-2}$ | |
| | $M$ (R ~ 100 000) | 4.8 | | 8 × 10$^{-18}$ erg s$^{-1}$ cm$^{-2}$ | |

Table 1. METIS filter bands and achievable sensitivities. The listed sensitivities of 0.48 μJy ($L'$), 4.8 μJy ($M'$), and 30 μJy ($N2$) correspond to $L'$ = 21.8 magnitudes, $M'$ = 19.0 magnitudes, and $N2$ = 15.1 magnitudes. The given ranges of spectroscopic sensitivities are caused by regions with high telluric line densities.

### Direct imaging in the $L$, $M$ and $N$ bands

Imaging in the $L$ and $M$ bands is performed with the short-wavelength arm (the $LM$-arm) of the imager, which provides a square FoV of 10.5 arcseconds on a side, sampling the focal plane with 5.5-milliarcsecond pixels. The long-wavelength arm (the $N$-arm) of the imager provides $N$-band imaging with a 13.5 × 13.5 arcseconds FoV and 6.8-milliarcsecond pixels. Observations are usually performed in pupil-tracking mode for optimal background rejection, with image de-rotation and stacking of the short exposures in post-processing. The expected sensitivities to point sources and spatially extended emission, as well as the available filter bands, are listed in Table 1.

### High-contrast imaging (HCI) in the $L$, $M$ and $N$ bands

METIS is equipped with several coronagraphs for HCI observations. The $LM$-arm of the imager contains an Apodised Phase Plate (APP), a coronagraph that is located in the pupil plane, works on all sources in the field, and is insensitive to pointing jitter. It delivers a dark region next to a bright source, where faint sources can be detected. The light from the bright source is not blocked. Both the $LM$-arm and the $N$-arm of the imager are equipped with several focal-plane-based coronagraphs. The Classical Vortex Coronagraph (CVC) efficiently blocks the light from a bright central source. It is only effective on-axis and requires the bright source to be precisely centred on the mask. Centring is achieved by applying the Quadrant Analysis of Coronagraph Images for Tip-Tilt Sensing (QACITS) algorithm on the science data at a closed loop control frequency of ~ 0.1 Hz. The vortex coronagraph can also be used in conjunction with an apodising pupil mask prior to the vortex phase mask, resulting in a Ring-Apodised Vortex Coronagraph (RAVC) for maximum contrast performance at the smallest angles from the central source.

### Long-slit spectroscopy in the $L$, $M$ and $N$ bands

Each arm of the METIS imager is equipped with grisms and a set of five 10-arcsecond-long slits, ranging in width from 19 to 114 milliarcseconds. Their spectral resolutions are listed in Table 1. For maximum efficiency, chopping and nodding are usually done in two or more positions along the slit. Long-slit observations will be performed in field-tracking mode to maintain the position angle of the slit on the sky. The telluric calibration is expected to rely primarily on fitting the observations to synthetic transmission spectra using Molecfit for a given atmospheric water vapour and temperature profile, measured by an on-site radiometer in the direction of observation.

### IFU spectroscopy in the $L$ and $M$ bands

The high-resolution (R ~ 100 000) spectrograph provides integral-field spectroscopy in the $L$ and $M$ bands. In its nominal mode the FoV of the IFU is 0.58 × 0.93 arcseconds, cut into 28 slices, each 0.021 arcseconds wide and 0.93 arcseconds long. All 28 slices of a selected spectral order are projected onto a 2 × 2 mosaic of detectors with a plate scale of 8.2 milliarcseconds per pixel. Since the point spread function is undersampled in the across-slice direction, a complete observation will consist of either a series of exposures with small dithers/offsets in the across-slice direction or a series of exposures where the field has been rotated by 90°, ensuring that both along- and across-slice directions are equally well sampled during part of that series. The telluric calibration is foreseen to be primarily model-based, analogues to the longslit modes.

The nominal simultaneous wavelength coverage (Table 1) can be increased in the so-called "extended wavelength coverage" mode, in which only three slices are projected on the focal plane, but with a larger $\Delta\lambda$ of the selected order than in the nominal mode. An instantaneous wavelength coverage of up to 300 nm can be achieved at the expense of spatial coverage.

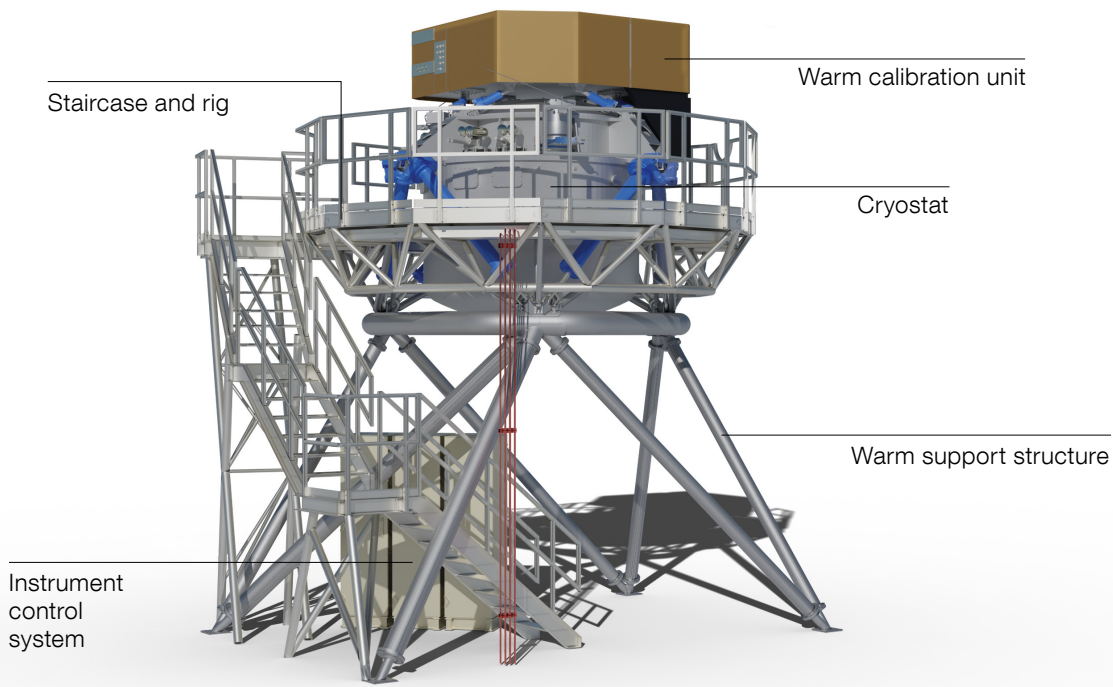

Figure 3. This engineering drawing of METIS shows its updated design following the Preliminary Design Review of the instrument.

### IFU spectroscopy combined with coronagraphy

The $LM$ high-resolution spectrograph can be used in combination with coronagraphy to suppress straylight from very bright sources. The observer can choose the APP coronagraph, which can be configured such that the point spread function dark region falls onto the image slicer while the bright source is directed toward a light trap. Alternatively, the observer can choose the Vortex Coronagraph, which rejects most of the starlight before encountering the IFU but allows a small fraction (≤ 10%) to be sent to the imager, where frames are being recorded for the QACITS algorithm to keep the source positioned accurately.

### Instrument performance

#### Image quality and contrast

METIS achieves fine-guiding accuracies below 0.02 $\lambda/D$ root-mean-square (RMS) on-sky, which corresponds to 0.4 milli-arcseconds RMS in the $L$ band, and 1 milliarcsecond RMS in the $N$ band. For bright ($K$ ≤ 10 mag) AO guide stars under median seeing conditions and moderate zenith angles (30 degrees), METIS will achieve a Strehl ratio of ≥ 87% at 3.7 μm (≥ 95% at 10 μm), which drops to ~ 62% (~ 93%) for a $K$ ~ 12 magnitude guide star. These numbers assume correction for non-common path aberrations and do not include degradation in the $N$ band due to water vapour seeing in conditions of high precipitable water vapour.

The excellent optical performance enables efficient coronagraphy. For the RAVC the 5σ contrast for a bright ($L$ ≤ 6 magnitudes) star in the L band after post-processing is estimated to be ~ 2 × 10$^{-4}$ at 2 $\lambda/D$ and ~ 2 × 10$^{-5}$ at 5 $\lambda/D$. The equivalent 5σ contrast for the APP is estimated to be ~ 2 × 10$^{-3}$ at 2 $\lambda/D$ and ~ 2 × 10$^{-5}$ at 5 $\lambda/D$.

#### Sensitivity

Table 1 lists the sensitivities METIS will achieve in the specified wavebands for both imaging and spectroscopy.

### References

Andrews, S. M. et al. 2018, ApJL, 869, L41
Anglada-Escudé, G. et al. 2016, Nature, 536, 437
Avenhaus, H. et al. 2018, ApJ, 863, 44
Banzatti, A. et al. 2017, ApJ, 834, 152
Birkby, J. L. et al. 2017, AJ, 153, 138
Brogi, M. et al. 2016, ApJ, 817, 106
Brown, J. M. et al. 2013, ApJ, 770, 94
Kasper, M. et al. 2019, The Messenger, 178, 5
Pascucci, I. et al. 2013, ApJ, 779, 178
Pinte, C. et al. 2018, ApJL, 860, L13
Quanz, S. P. et al. 2015, International Journal of Astrobiology, 14, 279
Smith, R. L. et al. 2009, ApJ, 701, 163
Teague, R. et al. 2018, ApJL, 860, L12

### Links

[1] METIS website: https://metis.strw.leidenuniv.nl/
[2] SimMETIS can be found at https://metis.strw.leidenuniv.nl/simmetis/
[3] The METIS app can be found at http://metis-app.strw.leidenuniv.nl/

### Notes

[a] The full list of METIS Consortium members can be found at https://metis.strw.leidenuniv.nl/about/